\documentclass[a4paper]{article}
\pdfoutput=1

\usepackage[utf8]{inputenc}
\usepackage[nocompress]{cite}
\usepackage{fullpage}
\usepackage{hyperref}
\usepackage{amsmath, amssymb}
\usepackage{graphicx}
\usepackage{amsthm} 
\usepackage{xcolor}

\newtheorem{theorem}{Theorem}

\newcommand{\Rank}{\textsf{rank}}
\newcommand{\Select}{\textsf{select}}
\newcommand{\Drank}{\textsf{subset-rank}}
\newcommand{\Dselect}{\textsf{subset-select}}

\date{}

\begin{document}

\title{Rank and Select on Degenerate Strings}

\author{%
Philip Bille\footnote{Supported by Danish Research Council grant DFF-8021-002498.}, Inge Li Gørtz\footnotemark[1], and Tord Stordalen\\[0.5em]
{\small\begin{minipage}{\linewidth}\begin{center}
\begin{tabular}{c}
DTU Compute, Technical University of Denmark\\
Lyngby, Denmark\\
\texttt{\{phbi, inge, tjost\}@dtu.dk}
\end{tabular}
\end{center}\end{minipage}}
}

\maketitle

\begin{abstract}
A \emph{degenerate string} is a sequence of subsets of some alphabet; it represents any string obtainable by selecting one character from each set from left to right. Recently, Alanko et al. generalized the rank-select problem to degenerate strings, where given a character \emph{c} and position \emph{i} the goal is to find either the \emph{i}th set containing \emph{c} or the number of occurrences of \emph{c} in the first \emph{i} sets~[SEA~2023]. The problem has applications to pangenomics; in another work by Alanko et al. they use it as the basis for a compact representation of \emph{de Bruijn Graphs} that supports fast membership queries. 

In this paper we revisit the rank-select problem on degenerate strings, introducing a new, natural parameter and reanalyzing existing reductions to rank-select on regular strings. Plugging in standard data structures, the time bounds for queries are improved exponentially while essentially matching, or improving, the space bounds.  Furthermore, we provide a lower bound on space that shows that the reductions lead to succinct data structures in a wide range of cases. Finally, we provide implementations; our most compact structure matches the space of the most compact structure of Alanko et al. while answering queries twice as fast. We also provide an implementation using modern vector processing features; it uses less than one percent more space than the most compact structure of Alanko et al. while supporting queries four to seven times faster, and has competitive query time with all the remaining structures. 

\end{abstract}

\section{Introduction}
Given a string $S$ over an alphabet $[1,\sigma]$ the \emph{rank-select problem} is to preprocess $S$ to support, for any $c \in [1,\sigma]$, 
\begin{itemize}
    \item $\Rank_S(i,c)$: return the number of occurrences of $c$ in $S[1,i]$
    \item $\Select_S(i,c)$: return the index $j$ of the $i$th occurrence of $c$ in $S$
\end{itemize}
This fundamental string problem has been studied extensively due to its wide applicability, see, e.g.,~\cite{BN15,OS07,GMR06,RRS07,PNB17,BCPT15,MN07,FMMN07,BHMS11,BCGNN14,NS14,HM10,NN14,GRRV13}, references therein, and surveys~\cite{Gagie16}.

A \emph{degenerate string} is a sequence $X = X_1,\ldots X_n$ where each $X_i$ is a subset of $[1,\sigma]$. We define its \emph{length} to be $n$, its \emph{size} to be $N = \sum_i |X_i|$, and denote by $n_0$ the number of empty sets among $X_1,\ldots,X_n$. Degenerate strings have been studied since the 80s~\cite{Abrahamson87} and the literature contains papers on problems such as degenerate string comparison~\cite{AABG+20}, finding string covers for degenerate strings\cite{CIKR+17}, and pattern matching with degenerate patterns, degenerate texts, or both~\cite{Abrahamson87, IMR08}. 

Alanko, Biagi, Puglisi, and Vuohtoniemi~\cite{ABPV23} recently generalized the rank-select problem to the \emph{subset rank-select problem}, where the goal is to preprocess a given degenerate string $X$ to support
\begin{itemize}
    \item $\Drank_X(i,c)$: return the number of sets in $X_1,\ldots,X_i$ that contain $c$
    \item $\Dselect_X(i,c)$: return the index of the $i$th set that contains $c$
\end{itemize}
Their motivation for studying this problem is to support fast membership queries on \emph{de Bruijn graphs}, a useful tool in pangenomic problems such as genome assembly and pangenomic read alignment (see~\cite{ABPV23,APV23} for details and further references). Specifically, in another work by some of the authors~\cite{APV23}, they show how to represent the de Bruijn graph of all length-$k$ substrings of a given string such that membership queries on the graph can be answered using $2k$ \Drank{} queries. They also provide an implementation that, when compared to the previous state of the art, improves query time by one order of magnitude while improving space usage, or by two orders of magnitude with similar space usage.

Their result for subset rank-select is the following~\cite{ABPV23}. They introduce the \emph{Subset Wavelet Tree}, a generalization of the well-known wavelet tree (see~\cite{GGV03}) to degenerate strings. It supports both \Drank{} and \Dselect{} queries in $O(\log \sigma)$ time and uses $2(\sigma - 1)n + o(n\sigma)$ bits of space in the general case. In the special case of  $n = N$ (which is the case for their representation of de Bruijn Graphs in~\cite{APV23}) they show that their structure uses $2n\log \sigma + o(n\log\sigma)$ bits. We note that their analysis for this special case happens to generalize nicely to also show that their structure uses at most  $2N\log\sigma + 2n_0 + o(N\log\sigma + n_0)$ bits for any $N$. 

Furthermore, in~\cite{APV23}, Alanko, Puglisi, and Vuohtoniemi present a number of reductions from the subset rank-select to the regular rank-select problem. We will elaborate on these reductions later in the paper.

\section{Our Results}

Our contributions are threefold. Firstly, we introduce the natural parameter $N$ and revisit the subset rank-select problem to reanalyze a number of simple and elegant reductions to the regular rank-select problem, based on the reductions from~\cite{APV23}. We express the complexities in terms of the performance of a given rank-select structure, achieving flexible bounds that benefit from the rich literature on the rank-select problem (Theorem~\ref{thm:general_upper_bound}). Secondly, we show that any structure supporting either $\Drank{}$ or $\Dselect{}$ must use at least $N\log \sigma - o(N\log \sigma)$ bits in the worst case (Theorem~\ref{thm:space_lower_bound}). By plugging a standard rank-select data structure into Theorem~\ref{thm:general_upper_bound} we, in many cases, match this bound to within lower order terms, while simultaneously matching the query time of the fastest known rank-select data structures (see below). Note that any lower bound for rank-select queries also holds for subset rank-select queries since any string is also a degenerate string. All our results hold on a word RAM with logarithmic word-size. Finally, we provide implementations of the reductions and compare them to the implementations of the Subset Wavelet Tree provided in~\cite{ABPV23}, and the implementations of the reductions provided in~\cite{APV23}. Our most compact structure matches the space of their most compact structure while answering queries twice as fast. We also provide a structure using vector processing features that matches the space of the most compact structure while imporving query time by a factor four to seven, remaining competitive with the fast structures for queries.

We now elaborate on the points above. The reductions are as follows.
\begin{theorem}\label{thm:general_upper_bound}
Let $X$ be a degenerate string of length $n$, size $N$, and with $n_0$ empty sets over an alphabet $[1,\sigma]$. Let $\mathcal{D}$ be a $\mathcal{D}_b(\ell, \sigma)$-bit data structure for a length-$\ell$ string over $[1,\sigma]$ that supports \Rank{} in $\mathcal{D}_r(\ell, \sigma)$ time and \Select{} in $\mathcal{D}_s(\ell, \sigma)$ time. If $n_0 = 0$ we can solve subset rank-select on $X$ in\\
(i) $\mathcal{D}_b(N,\sigma) + N + o(N)$ bits, $\mathcal{D}_r(N,\sigma) + O(1)$ \Drank{}-time, and $\mathcal{D}_s(N,\sigma) + O(1)$ \Dselect{}-time.\\
Otherwise, if $n_0 > 0$ we can solve subset rank-select on $X$ in\\
(ii) the bounds in (i) where we replace $N$ by $N' = N + n_0$ and $\sigma$ by $\sigma' = \sigma + 1$. \\
(iii) the bounds in (i) with additional $\mathcal{B}_b(n, n_0)$ bits of space, additional $\mathcal{B}_r(n,n_0)$ time for \Drank{}, and additional $\mathcal{B}_s(n,n_0)$ time for \Dselect{}. Here $\mathcal{B}$ is a data structure on a length-$n$ bitstring that contains $n_0$ \texttt{1}s, uses $\mathcal{B}_b(n,n_0)$ bits, and supports $\Rank{}(\cdot, \texttt{1})$ in $\mathcal{B}_r(n,n_0)$ time and $\Select{}(\cdot,\texttt{0})$ in $\mathcal{B}_s(n,n_0)$ time. 
\end{theorem}

Here Theorem~\ref{thm:general_upper_bound}(i) and~(ii) are based on the reduction from~\cite[Sec. 4.3]{APV23}, and Theorem~\ref{thm:general_upper_bound}(iii) is a variation of Theorem~\ref{thm:general_upper_bound}(ii) that handles empty sets using a natural, alternative strategy. By plugging a standard rank-select structure into Theorem~\ref{thm:general_upper_bound} we exponentially improve query times while essentially matching, or improving, space usage compared to Alanko et al.~\cite{ABPV23}. For example, consider the rank-select structure by Golynski, Munro, and Rao~\cite{GMR06} which uses $\ell\log\sigma + o(\ell\log\sigma)$ bits, supports \Rank{} in $O(\log\log \sigma)$ time, and supports \Select{} in constant time. These query times are optimal in succinct space, see e.g.~\cite{BN15}.
        
For $n_0 = 0$, plugging this structure into Theorem~\ref{thm:general_upper_bound}(i) yields an $N\log \sigma + N +  o(N\log \sigma + N)$  bit data structure supporting \Drank{} in $O(\log\log \sigma)$ time and \Dselect{} in constant time. Compared to the previous result by Alanko et al.~\cite{ABPV23}, this improves the constant on the space bound from $2$ to $1 + 1/\log \sigma$ and improves the query time from $O(\log \sigma)$ for both queries to $O(\log\log \sigma)$ for \Drank{} and constant for \Dselect{}.  Note that the additional $N$ bits in the space bound are a lower order term when $\sigma = \omega(1)$. 

For $n_0 > 0$, plugging their structure into Theorem~\ref{thm:general_upper_bound}(ii) gives the same time bounds as above and the space bound 
\[
    (N + n_0)\log(\sigma + 1) + (N + n_0) + o(n_0\log\sigma + N\log\sigma + N + n_0)
\]        
bits. If $n_0 = o(N)$ and $\sigma = \omega(1)$, the space bound is identical to the one above. In any case, the query time is still improved exponentially.

Alternatively, by plugging it into Theorem~\ref{thm:general_upper_bound}(iii) the space bound becomes $N\log\sigma + o(N\log\sigma)$ + $\mathcal{B}_s(n, n_0)$ bits. For $n = o(N\log\sigma)$ we can choose $\mathcal{B}$ to be an $(n + o(n))$-bit data structure with constant time \Rank{} and \Select{}, such as~\cite{CM96,Jacobson89}, again achieving the same space and time bounds as when $n_0 = 0$. Otherwise, we can plug in any data structure for $\mathcal{B}$ that is sensitive to the number of $1$-bits in the bitvector. For example, if $n_0 = O(\log n)$ we can store the positions of the $1$-bits in sorted order using $O(n_0\log n) = O(\log^2 n)$ bits, supporting $\Select{}(i,\texttt{1})$ in constant time and $\Rank{}(i,\cdot)$ in $O(\log n_0) = O(\log\log n)$ time using binary search. 
We can also binary search for $\Select{}(i,\texttt{0})$ in $O(\log n_0) = O(\log\log n)$ time using the fact that --- if the $i$th position of a $1$-bit is $p_i$ --- there are $p_i - i$ zeroes in the prefix ending at $p_i$. There are many such sensitive data structures that obtain various time-space trade-offs, e.g~\cite{OS07,GORR14}.

We also show the following lower bound on the space required to support either \Drank{} or \Dselect{} on a degenerate string. 
\begin{theorem}\label{thm:space_lower_bound}
Let $X$ be a degenerate string of size $N$ over an alphabet $[1,\sigma]$. Any data structure supporting \Drank{} or \Dselect{} on $X$ must use at least $N\log \sigma - o(N\log \sigma)$ bits in the worst case.             
\end{theorem}
Thus, applying Theorem~\ref{thm:general_upper_bound} in many cases results in \textit{succinct} data structures, whose space deviates from this lower bound by at most a lower order term. The three examples above each illustrate this when respectively  (1) $\sigma = \omega(1)$, (2) $n_0 = o(N)$ and $\sigma = \omega(1)$, and (3) $n = o(N\log \sigma)$. 

Finally, we provide implementations and compare them to variants of the Subset Wavelet Tree~\cite{ABPV23} and the reductions~\cite{APV23} implemented by Alanko et al. Specifically, we apply the test framework from~\cite{ABPV23} and run two types of tests: one where the subset rank-select structures are used to support $k$-mer queries on a de Bruijn Graph (the motivation for,  and practical application of, the subset rank-select problem), and one where \Drank{} queries are tested in isolation. We implement Theorem~\ref{thm:general_upper_bound}(iii) and plug in efficient off-the-shelf rank-select structures from the \emph{Succinct Data Structure Library (SDSL)}\footnote{\small \url{https://github.com/simongog/sdsl-lite}}~\cite{GBMP14}. We also implement a variation of another reduction from~\cite[Sec. 4.2]{APV23}, which is more optimized for genomic test data. The highlight is our most compact structure, which matches the space of their most compact structure while supporting queries twice as fast, as well as our structure using vector processing, which matches the most compact structure while supporting queries four to seven times faster.

\section{Reductions}{\label{sec:reductions}}
We now present the reductions from Theorem~\ref{thm:general_upper_bound}. Let $X$, $\mathcal{D}$, and $\mathcal{B}$ be defined as in Theorem~\ref{thm:general_upper_bound}. Furthermore, let $\mathcal{V}$ be the data structure from~\cite{Jacobson89}, which for a length-$\ell$ bitstring uses $\ell + o(\ell)$ bits and supports \Rank{} and \Select{} in constant time.

\subsection{Reductions~(i) and~(ii)}
First assume that $n_0 = 0$. For each $X_i$ let the string $S_i$ be the concatenation of the characters in $X_i$ in an arbitrary order, and let the string $R_i$ be a single \texttt{1} followed by $|X_i| - 1$ \texttt{0}s. This is always possible since $|X_i| \geq 1$. Let $S$ (resp. $R$) be the concatenation of $S_1,\ldots,S_n$ (resp. $R_1,\ldots,R_n)$ in that order, with an additional \texttt{1} appended after $R_n$. The lengths of $S$ and $R$ are respectively $N$ and $N+1$. See Figure~\ref{fig:reduction_example} for an example. The data structure consists of $\mathcal{D}$ built over $S$ and $\mathcal{V}$ built over $R$, which takes $\mathcal{D}(N,\sigma) + N + o(N)$ bits.

To support $\Drank{}(i,c)$, compute the starting position $k = \Select{}_R(i+1,\texttt{1})$ of $S_{i+1}$ and return $\Rank{}_S(k-1, c)$. To support $\Dselect{}(i,c)$, find the index $k = \Select{}_S(i,c)$ of the $i$th occurrence of $c$, and return $\Rank{}_R(k,\texttt{1})$ to determine which set $k$ is in. Since \Rank{} and \Select{} queries on $R$ take constant time, $\Drank{}$ and $\Dselect{}$ queries take respectively $\mathcal{D}_r(N,\sigma) + O(1)$ and $\mathcal{D}_s(N,\sigma) + O(1)$ time, achieving the bounds stated in Theorem~\ref{thm:general_upper_bound}(i).

If $n_0 \neq 0$, add a new character $\sigma + 1$ and replace each empty set with the singleton set $\{\sigma + 1\}$, and then apply reduction~(i). This instance has $N' = N+n_0$ and $\sigma' = \sigma + 1$, achieving the bounds in Theorem~\ref{thm:general_upper_bound}(ii).

\subsection{Reduction~(iii)}
Let $E$ denote the length-$n$ bitvector where $E[i] = 1$ if $X_i = \emptyset$ and $E[i] = 0$ otherwise. Let $X''$ denote the degenerate string obtained by removing all the empty sets from $X$. The data structure consists of reduction~(i) over $X''$ and $\mathcal{B}$ built over $E$. This takes $\mathcal{D}_b(N,\sigma) + N + o(N) + \mathcal{B}_b(n, n_0)$ bits. To support $\Drank{}_X(i,c)$ first compute $k = i - \Rank{}_E(i,\texttt{1})$, mapping $X_i$ to its corresponding set $X''_{k}$. Then return $\Drank{}_{X''}(k,c)$. This takes $\mathcal{B}_r(n,n_0) + \mathcal{D}_r(N,\sigma) + O(1)$ time. 
To support $\Dselect{}_X{}(i,c)$, find $k = \Dselect{}_{X''}(i,c)$ and return $\Select{}_E(k, \texttt{0})$, the position of the $k$th zero in $E$ (i.e., the $k$th non-empty set). This takes $\mathcal{B}_s(n,n_0) + \mathcal{D}_s(N,\sigma) + O(1)$, matching the stated bounds.

        \begin{figure}
            \centering
            
            \begin{tabular}{c c c c c }$X = $ & 
                $\Bigg\{\,\begin{matrix}\texttt{A}\\\texttt{C}\\\texttt{G}\end{matrix}\,\Bigg\}$ &
                $\Bigg\{\,\begin{matrix}\texttt{A}\\\texttt{T}\end{matrix}\,\Bigg\}$ &
                $\Bigg\{\,\begin{matrix}\texttt{C}\end{matrix}\,\Bigg\}$ &
                $\Bigg\{\,\begin{matrix}\texttt{T}\\\texttt{G}\end{matrix}\,\Bigg\}$  \\ & & & & \\
                & $X_1$ & $X_2$ & $X_3$ & $X_4$ 
            \end{tabular}\qquad\begin{tabular}{c c c  c c c}
                $S =$ & \texttt{ACG} & \texttt{AT} &  \texttt{C} & \texttt{TG} & \\
                $R = $ & \texttt{100} & \texttt{10}  & \texttt{1} & \texttt{10} & \texttt{1}\\\\
                & $S_1$ & $S_2$ & $S_3$ & $S_4$
                
            \end{tabular}
            \caption{\emph{Left:} A degenerate string $X$ over the alphabet $\{\texttt{A}, \texttt{C}, \texttt{G}, \texttt{T}\}$ where $n = 4$ and $N = 8$. \emph{Right:} The reduction from Theorem~\ref{thm:general_upper_bound}(i) on $X$. White space is for illustration purposes only. To compute $\Drank{}(2,\texttt{A})$, we first compute $\Select{}_R(3,\texttt{1}) = 6$. Now we know that $S_2$ ends at position $5$, so we return $\Rank{}_S(5, \texttt{A}) = 2$. To compute $\Dselect{}(2, \texttt{G})$ we compute $\Select{}_S(2,\texttt{G}) = 8$, and compute $\Rank{}_R(8,\texttt{1}) = 4$ to determine that position 8 corresponds to $X_4$.}
            \label{fig:reduction_example}
        \end{figure}

\section{Lower Bound}\label{sec:lower_bound}
In this section we prove Theorem~\ref{thm:space_lower_bound}. The strategy is as follows. Any structure supporting $\Drank{}$ or $\Dselect{}$ on $X$ is a representation of $X$ since we can fully recover $X$ by repeatedly using either of these operations. We will lower bound the number $L$ of distinct degenerate strings that can exist for a given  $N$ and $\sigma$. Any representation of $X$ must be able to distinguish between these instances, so it needs to use at least $\log_2 L$ bits in the worst case. Let sufficiently large $N$ and $\sigma = \omega(\log N)$ be given and assume without loss of generality that $\log N$ and $N/\log N$ are integers. Consider the class of degenerate strings $X_1,\ldots,X_n$ where each $|X_i| = \log N$ and $n = N/\log N$. There are
$
    \binom{\sigma}{\log N}^{N/\log N} 
$
such degenerate strings, so any representation must use at least
        \begin{align*}
            \log \binom{\sigma}{\log N}^{N/\log N} &= \frac{N}{\log N}\log \binom{\sigma}{\log N}\\ &\geq \frac{N}{\log N}\log 
              \left(\frac{\sigma - \log N}{\log N}\right)^{\log N}\\
              &= N\log \left(\frac{\sigma - \log N}{\log N}\right)\\
              &= N\log \sigma - o(N\log \sigma)
        \end{align*}
        bits, concluding the proof.    

\section{Experimental Setup}\label{sec:experiments}
\subsection{Setup and Data}
The code to replicate our results is available on GitHub\footnote{\small \url{https://github.com/tstordalen/subset-rank-select}}. Our tests are based on the test framework by Alanko et al.~\cite{ABPV23}, also available on GitHub\footnote{\small \url{https://github.com/jnalanko/SubsetWT-Experiments/}}. Like them, we used the following data sets. 
    \begin{enumerate}
        \item A pangenome of $3682$ E. coli genomes, available on Zenodo\footnote{\small \url{https://zenodo.org/record/6577997}}. According to~\cite{ABPV23}, the data was collected by downloading a set of 3682 E. Coli assemblies from the National Center for Biotechnology Information. 
        \item A human metagenome (SRA identifier ERR$5035349$) consisting of a set of $\approx 17$ million length-$502$ sequence reads sampled from the human gut  from a study on irritable bowel syndrome and bile acid malabsorption~\cite{JDOE+20}.  
    \end{enumerate}
We applied two tests. Firstly, we plugged our data structures into the $k$-mer query test from~\cite{ABPV23}; they plug subset rank-select structures into their $k$-mer index and query a large number of $k$-mers. Secondly, we tested the subset rank-select structures in isolation by building the $k$-mer indices, extracting the subset rank-select structures, and performing twenty million randomly generated $\Drank$ queries. For each measurement we built only the structure under testing, and timed only the execution of the queries. Each value reported below is the average of five such measurements. Note that, like~\cite{ABPV23}, we do not test \Dselect{} queries; only \Drank{} queries are necessary for their $k$-mer index. 

All the tests were run on a system with a $3.00$GHz i7-1185G7 processor and $32$ gigabytes of DDR4 random access memory, running Ubuntu $22.04.3$ LTS with kernel version 6.2.0-35-generic. The programs were compiled using g++ version 11.4.0 with compiler flags \texttt{-O3}, \texttt{-march=native}, and \texttt{-DNDEBUG}. 

\subsection{Data Structures}\label{sec:data_structures}
This section summarizes a representative subset of the data structures we tested; see appendix~\ref{appendix:extra_data_structures} for a description of, and results for, the remaining data structures. We implement both Theorem~\ref{thm:general_upper_bound}(iii) as well as variation of the reduction \emph{split representation} from~\cite[Sec 4.2]{APV23}; this reduction is optimized for their $k$-mer query structure built over genomic data, in which most of the sets are singletons. We name our variation the \emph{dense-sparse decomposition (DSD)}, which works as follows. The empty sets are handled in the same way as in Theorem~\ref{thm:general_upper_bound}(iii). Furthermore, we store a sparse bitvector of length $n$ for each character, i.e., \texttt{A}, \texttt{C}, \texttt{G}, and \texttt{T}. For each $X_i$ of size at least two we remove $|X_i| - 1$ of the characters and set the $i$th bit in the corresponding bitvector to $1$. What remains are $n - n_0$ singleton sets, i.e., a regular string, for which we store a rank-select structure. A query thus consists of three rank queries; one to eliminate empty sets, one in the regular string, and one in the sparse bitvector. In the split representation by~\cite{APV23}, each such set is instead removed and \emph{all} the characters are represented in the additional bitvectors. 

The data structures we tested are as follows. \textbf{Matrix} is the benchmark structure from~\cite{ABPV23}, consisting of one bitvector per character (i.e., a $4\times n$ matrix). \textbf{Thm~\ref{thm:general_upper_bound}(iii)} is the reduction from Theorem~\ref{thm:general_upper_bound}(iii), using a wavelet tree for the string, a bitvector for the length-$N$ indicator string, and a sparse bitvector for the empty sets. \textbf{DSD~(x)}, \textbf{SWT~(x)}, and \textbf{Split~(x)} are the DSD, Subset Wavelet Tree, and split representation parameterized by \textbf{x}, respectively, where \textbf{x} may be any of the following data structures:  \emph{(1)}  \textbf{scan}, the structure from Alanko et al.~\cite[Sec. 5.2]{ABPV23}, inspired by scanning techniques for fast \Rank{} queries on bitvectors, \emph{(2)} \textbf{split}, a rank structure for size-four alphabets optimized for the skewed distribution of singleton to non-singleton sets~\cite[Sec 5.3]{ABPV23} (not to be confused with the split representation) \emph{(3)} \textbf{rrr}, an SDSL wavelet tree using $H_0$-compressed bitvectors, based mainly on the result by Raman, Raman, and Rao Satti~\cite{RRS07}, \emph{(4)} \textbf{rrr gen.}, a generalization of RRR to size-four alphabets~\cite[Sec. 5.4]{ABPV23}, \emph{(5)} \textbf{ef}, an efficient implementation of rank queries on a bitvector stored using Elias-Fano encoding from~\cite{MPRZ21}, and \emph{(6)} \textbf{plain}, a standard SDSL bitvectors supporting rank in constant time. 

Furthermore,~\cite{APV23} implements Concat~(rrr), which is essentially reduction~(ii) using a wavelet tree with RRR-compressed bitvectors, and we also implement the structure DSD (SIMD). It is based on a standard idea for compact data structures; we divide the string into blocks, precompute the answer to rank queries up to each block, and compute partial rank queries for blocks as needed using word parallelism (this is also an essential idea in the `scan' structure by~\cite{ABPV23}). We use \emph{SIMD (single instruction, multiple data)} instructions to speed up the partial in-block rank queries, which allows for large blocks and a reduction in space (see Appendix~\ref{appendix:extra_data_structures}). Most computers support SIMD to some extent, allowing the same operation to be performed on many words simultaneously. We used AVX512, which supports $512$-bit vector registers.

\begin{figure}[t]
    \centering
    \includegraphics{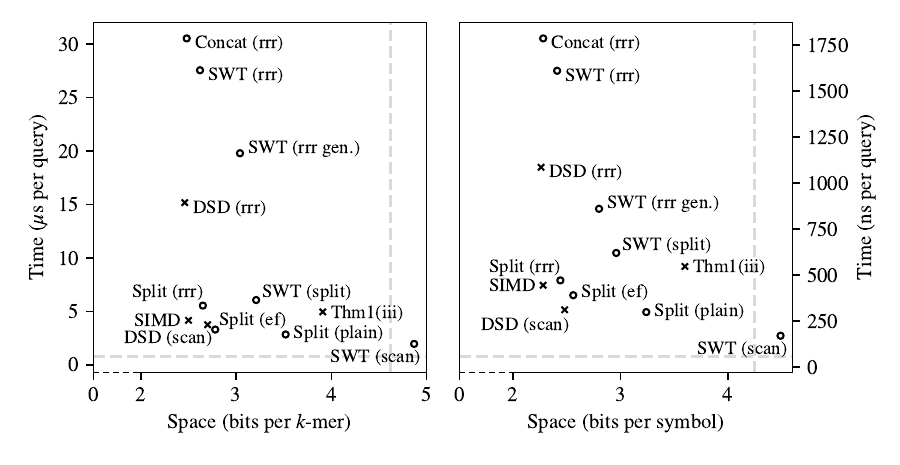}
    \caption{Note that the $x$-axis is truncated in both plots. The two gray lines represent the performance of the benchmark solution ``Matrix''. The crosses indicate our data structures and the circles indicate the data structures from~\cite{ABPV23, APV23}.  \emph{Left:} Results of the $k$-mer query test on the metagenome data set. \emph{Right:} The result of the \Drank{} test on the metagenome data set. The space is in number of bits per symbol, i.e., $\text{bits}/N$.}
    \label{fig:metagenome_results}
\end{figure}

\section{Results}\label{sec:results}
The test results for the metagenome data set can be seen in Figure~\ref{fig:metagenome_results}; the results for the E. Coli data set are similar. See appendix~\ref{appendix:data} for the data belonging to Figure~\ref{fig:metagenome_results}, and appendix~\ref{appendix:extra_data_structures} for the results of the data structures omitted from this figure. The fastest structure is SWT (scan), but it is large and is outperformed by the benchmark solution on both parameters. Our unoptimized reduction Thm1(iii) uses $20-60\%$ more space than the remaining structures of~\cite{ABPV23,APV23} while remaining within a factor two in query time of most of them. Our fastest structure, DSD~(scan), is competitive with both Split~(ef) and Split~(rrr). Our most compact structure DSD~(rrr) matches the space of the previous smallest structure, Concat~(ef), while supporting queries twice as fast. Our SIMD-enhanced structure uses less than one percent more space than Concat~(ef) while supporting queries four to seven times faster. It is also competitive with the fast and compact structures Split~(ef) and Split~(rrr). We note that the entropies for the distributions of sets in the Metagenome and E. Coli data sets are respectively $2.21$ and $2.24$ bits (as seen in~\cite{ABPV23}), and that reduction from $2.44$ bits (Split~(rrr), Metagenome) to $2.28$ bits (SIMD, Metagenome) reduces the distance to the entropy from approximately $10\%$ to $3\%$, while simultaneously supporting queries faster.

\bibliographystyle{plainurl}
\bibliography{refs}

\newpage
\appendix
\section{Additional Data}\label{appendix:data}
\begin{table}[h]
    \renewcommand{\arraystretch}{1.5}
    \centering
    \begin{tabular}{l | c c | c c || c c | c c |}
         & \multicolumn{4}{c||}{$k$-mer Queries} & \multicolumn{4}{c|}{Subset Rank Queries}\\
         & \multicolumn{2}{c}{E. Coli} & \multicolumn{2}{|c||}{Metagenome} & \multicolumn{2}{c}{E. Coli} & \multicolumn{2}{|c|}{Metagenome}\\
         Data structure & {\small Query} & {\small Space} & {\small Query} & {\small Space} & {\small Query} & {\small Space} & {\small Query} & {\small Space}  \\ 
          &  ($\mu$s)  & (bpk)  &  ($\mu$s)  & (bpk)  &  (ns) & (bps) & (ns) & (bps) \\\hline
        Matrix & 0.63 & 4.29 & 0.77 & 4.62 & 38.75 & 4.26 & 56.98 & 4.25 \\\hline
        DSD (scan) & 3.00 & 2.61 & 3.75 & 2.70 & 210.23 & 2.57 & 311.33 & 2.48 \\
        Thm1(iii) & 3.87 & 3.68 & 4.95 & 3.91 & 435.28 & 3.64 & 546.89 & 3.60 \\
        DSD (rrr) & 13.21 & 2.38 & 15.17 & 2.46 & 850.99 & 2.34 & 1086.11 & 2.26 \\\hline
        SIMD & 3.31 & 2.42 & 4.16 & 2.50 & 320.53 & 2.37 & 444.94 & 2.28 \\\hline
        SWT (scan) & 1.63 & 4.53 & 1.96 & 4.87 & 129.44 & 4.49 & 170.44 & 4.49 \\
        SWT (split) & 4.93 & 3.17 & 6.06 & 3.21 & 436.69 & 3.13 & 620.47 & 2.96 \\
        SWT (rrr gen.) & 18.97 & 2.84 & 19.79 & 3.04 & 789.12 & 2.81 & 860.4 & 2.80 \\
        SWT (rrr) & 25.33 & 2.48 & 27.55 & 2.62 & 1384.0 & 2.45 & 1610.73 & 2.41 \\\hline
        Split (plain) & 2.28 & 3.30 & 2.84 & 3.52 & 235.22 & 3.27 & 298.87 & 3.24 \\
        Split (ef) & 2.71 & 2.69 & 3.30 & 2.78 & 317.71 & 2.65 & 390.65 & 2.56 \\
        Split (rrr) & 4.70 & 2.54 & 5.54 & 2.65 & 393.14 & 2.51 & 471.30 & 2.44 \\
        Concat(ef) & 26.25 & 2.38 & 30.53 & 2.48 & 1372.2 & 2.35 & 1786.65 & 2.28 \\\hline
    \end{tabular}
    \caption{The left half of the table shows the result for the $k$-mer query test. The times are listed in microseconds per query, and space in the number of bits per represented $k$-mer. The right half shows the result of the \Drank{} query test. Times are listed in nanoseconds per query, and space in bits per symbol (i.e., the number of bits divided by $N$). There are five groups of data structures, separated by horizontal lines; the benchmark structure, our reductions, our structure using SIMD, the Subset Wavelet Trees from~\cite{ABPV23}, and the reductions from~\cite{APV23}.  Each group is ordered from fastest to slowest and largest to smallest, except for Thm1(iii) which breaks space order. Each value in the table is the average of five measurements.}
    \label{tab:kmer_data}
\end{table}

\section{Results for all Data Structures}\label{appendix:extra_data_structures}
This section elaborates on the data structures that were omitted from Sections~\ref{sec:data_structures} and~\ref{sec:results}. From~\cite{APV23}, we omitted the structure \textbf{Concat~(plain)}, which is the structure \textbf{Concat~(ef)} explained in Section~\ref{sec:data_structures} parameterized with standard SDSL bitvcectors instead. We also omitted \textbf{Matrix~(ef)} and \textbf{Matrix~(rrr)}, which is the benchmark solution parameterized with different types of bitvectors.

Furthermore, this section includes more parameterizations of our SIMD enhnaced data structure. We elaborate on the SIMD structure from Section~\ref{sec:data_structures}. \textbf{DSD~(SIMD~(i))} (which we refer to as SIMD~(i) for simplicity) is the DSD parameterized with the SIMD~(i) structure, which supports rank queries on strings over the alphabet $\{0,1,2,3\}$. As described in Section~\ref{sec:data_structures}, the main idea is to divide the string into blocks, precompute the answer to rank queries up to the start of each block, and compute partial rank queries internally in blocks as needed using SIMD. The width of the block is determined by the parameter $i$; there are $512i$ characters stored per block (recall that the width of the SIMD registers we used was $512$ bits). The blocks are stored as follows. Each character in $\{0,1,2,3\}$ consists of two bits. We separate the string represented by the block into two bitstrings; one consisting of the low bits, and one of the high bits. To compute partial rank queries, we use the operation \texttt{vpternlogq}, which given three vectors and an $8$-bit integer value evaluates \emph{any} three-variable boolean function bit-wise for the three vectors (the $8$-bit integer describes the result column of the three-variable truth table, which has eight rows). The operation also accepts an additional integer, used to mask out results if they are not needed for the computation. To answer rank queries, we traverse the low and high bitstrings, using \texttt{vpternlogq} to find occurrences of the queried character and the mask to filter out results occurring after the queried index. The version of the SIMD data structure in Sections~\ref{sec:data_structures} and~\ref{sec:results} is SIMD~(8).

\begin{figure}[h]
    \centering
    \includegraphics[width=1\textwidth]{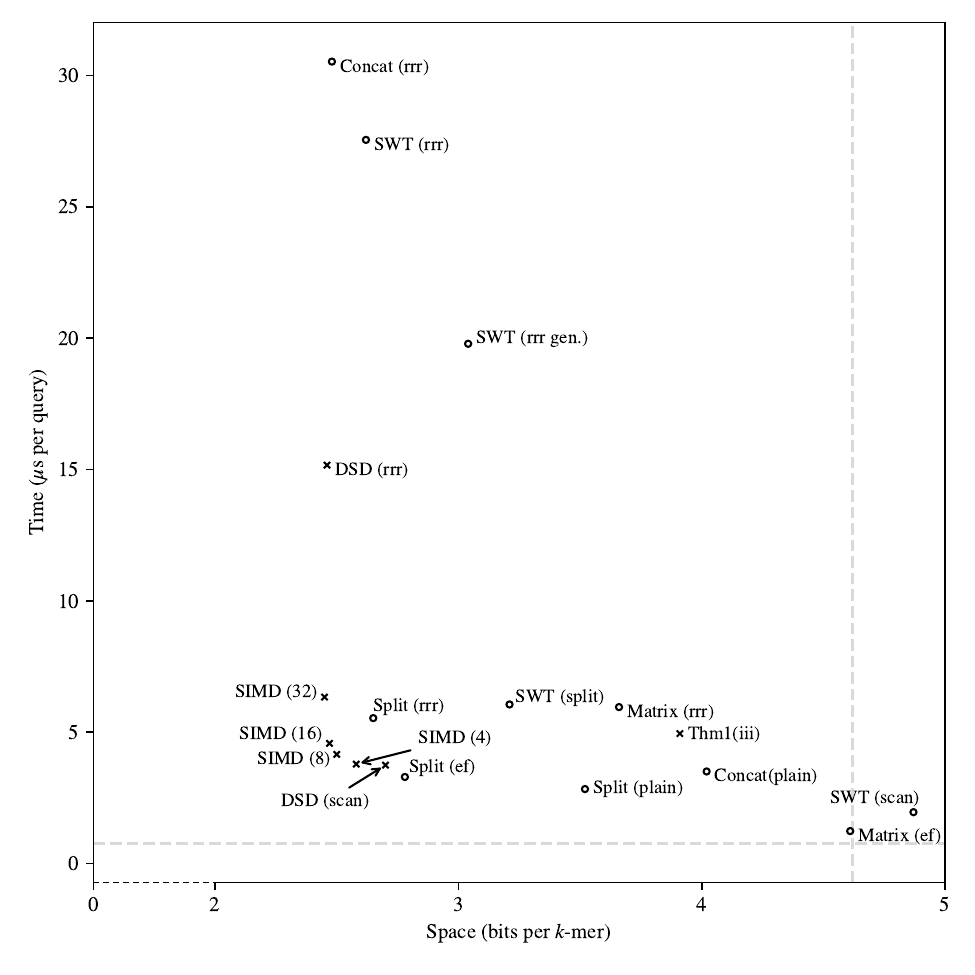}
    \caption{Note that the $x$-axis is truncated. Shows the performance of all data structures for the $k$-mer search test on the metagenome data set. The two gray lines represent the performance of the benchmark solution ``Matrix''. The crosses indicate our data structures and the circles indicate the data structures from~\cite{ABPV23, APV23}.}
    \label{fig:all_structures_kmer_search}
\end{figure}

\begin{figure}[h]
    \centering
    \includegraphics[width=1\textwidth]{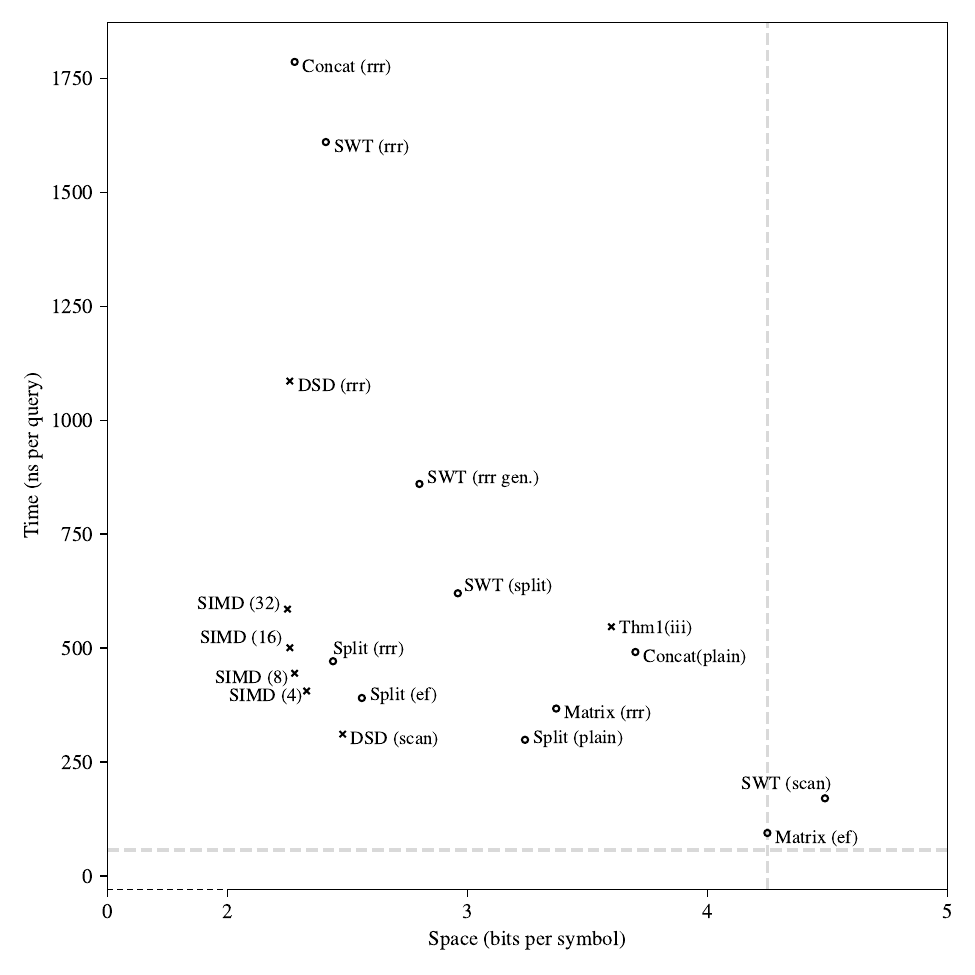}
    \caption{Note that the $x$-axis is truncated. Shows the performance of all data structures for the rank query benchmark. The space is in number of bits per symbol, i.e., $\text{bits}/N$.  The two gray lines represent the performance of the benchmark solution ``Matrix''. The crosses indicate our data structures and the circles indicate the data structures from~\cite{ABPV23, APV23}.}
    \label{fig:all_structures_rank_becnhmark}
\end{figure}

The results for all the structures on the metagenome data set can be seen in Figure~\ref{fig:all_structures_kmer_search} ($k$-mer search test) and Figure~\ref{fig:all_structures_rank_becnhmark} (rank query benchmark). All the data, for all structures, both data sets, and both tests, can be seen in Table~\ref{tab:all_structures_data}.

\begin{table}[h]
    \renewcommand{\arraystretch}{1.5}
    \centering
    \begin{tabular}{l | c c | c c || c c | c c |}
         & \multicolumn{4}{c||}{$k$-mer Queries} & \multicolumn{4}{c|}{Subset Rank Queries}\\
         & \multicolumn{2}{c}{E. Coli} & \multicolumn{2}{|c||}{Metagenome} & \multicolumn{2}{c}{E. Coli} & \multicolumn{2}{|c|}{Metagenome}\\
         Data structure & {\small Query} & {\small Space} & {\small Query} & {\small Space} & {\small Query} & {\small Space} & {\small Query} & {\small Space}  \\ 
          &  ($\mu$s)  & (bpk)  &  ($\mu$s)  & (bpk)  &  (ns) & (bps) & (ns) & (bps) \\\hline
        Matrix       & 0.63 & 4.29 & 0.77 & 4.62 & 38.75 & 4.26 & 56.98 & 4.25 \\
        Matrix (ef)  & 0.92 & 4.07 & 1.24 & 4.61 & 73.9 & 4.04 & 94.4 & 4.25 \\
        Matrix (rrr) & 5.28 & 3.38 & 5.96 & 3.66 & 301.7 & 3.34 & 367.45 & 3.37 \\
        \hline
        DSD (scan) & 3.0 & 2.61 & 3.75 & 2.7 & 210.23 & 2.57 & 311.33 & 2.48 \\
        Thm1(iii) & 3.87 & 3.68 & 4.95 & 3.91 & 435.28 & 3.64 & 546.89 & 3.6 \\
        DSD (rrr) & 13.21 & 2.38 & 15.17 & 2.46 & 850.99 & 2.34 & 1086.11 & 2.26 \\\hline
        SIMD (4) & 2.98 & 2.49 & 3.79 & 2.58 & 289.54 & 2.42 & 405.88 & 2.33 \\
        SIMD (8) & 3.31 & 2.42 & 4.16 & 2.5 & 320.53 & 2.37 & 444.94 & 2.28 \\
        SIMD (16) & 3.67 & 2.39 & 4.58 & 2.47 & 371.87 & 2.35 & 500.97 & 2.26 \\
        SIMD (32) & 5.35 & 2.37 & 6.34 & 2.45 & 447.28 & 2.34 & 585.65 & 2.25 \\\hline
        SWT (scan) & 1.63 & 4.53 & 1.96 & 4.87 & 129.44 & 4.49 & 170.44 & 4.49 \\
        SWT (split) & 4.93 & 3.17 & 6.06 & 3.21 & 436.69 & 3.13 & 620.47 & 2.96 \\
        SWT (rrr gen.) & 18.97 & 2.84 & 19.79 & 3.04 & 789.12 & 2.81 & 860.4 & 2.8 \\
        SWT (rrr) & 25.33 & 2.48 & 27.55 & 2.62 & 1384.0 & 2.45 & 1610.73 & 2.41 \\\hline
Concat(plain) & 2.53 & 3.74 & 3.51 & 4.02 & 375.44 & 3.71 & 491.55 & 3.7 \\
Concat(ef) & 26.25 & 2.38 & 30.53 & 2.48 & 1372.2 & 2.35 & 1786.65 & 2.28 \\\hline
    Split (plain) & 2.28 & 3.3 & 2.84 & 3.52 & 235.22 & 3.27 & 298.87 & 3.24 \\
    Split (ef) & 2.71 & 2.69 & 3.3 & 2.78 & 317.71 & 2.65 & 390.65 & 2.56 \\
    Split (rrr) & 4.7 & 2.54 & 5.54 & 2.65 & 393.14 & 2.51 & 471.3 & 2.44 \\\hline
    
    \end{tabular}
    \caption{The left half of the table shows the result for the $k$-mer query test. The times are listed in microseconds per query, and space in the number of bits per represented $k$-mer. The right half shows the result of the \Drank{} query test. Times are listed in nanoseconds per query, and space in bits per symbol (i.e., the number of bits divided by $N$). There are six groups of data structures, separated by horizontal lines; the variants of the benchmark structure, our reductions, our structure using SIMD, the Subset Wavelet Trees from~\cite{ABPV23}, the ``Concat'' reduction from~\cite{APV23}, and the ``Split'' reduction from~\cite{APV23}. Each group is ordered from fastest to slowest and largest to smallest, except for Thm1(iii) which breaks space order. Each value in the table is the average of five measurements.}
    \label{tab:all_structures_data}
\end{table}

\end{document}